\documentclass[12pt,english]{article}
\usepackage{babel}
\usepackage{amsfonts,epsfig}
\title{\bf Non-perturbative solutions in the electro-weak theory with $\bar t t$ condensate and the $t$-quark mass}
\author{{\bf Boris A. Arbuzov\thanks{E-mail: arbuzov@theory.sinp.msu.ru}  and Ivan V. Zaitsev}\\
{Skobeltsyn Institute of Nuclear Physics of
MSU,}\\ {119991 Moscow, Russia}
}

\date{}
\begin{document}
\maketitle
\begin{quote}
We apply Bogoliubov compensation principle to the gauge electro-weak interaction
to demonstrate a  spontaneous generation of anomalous three-boson gauge invariant
effective interaction. The non-trivial solution of compensation equations uniquely defines the form-factor of the anomalous interaction and parameters of the theory including value of gauge electro-weak coupling $g(M_W^2)\simeq 0.62$ in satisfactory agreement to the experimental value. 
A possibility of spontaneous generation of effective four-fermion interaction of heavy quarks is also demonstrated.  
This interaction defines an equation for a scalar bound state of heavy quarks which serve as a substitute for the elementary scalar Higgs doublet. As a result we calculate the $t$-quark mass $m_t\,=\,177\,GeV$ in satisfactory agreement with the experimental value. The results strongly support idea of $\bar t\,t$ condensate as a source of the electro-weak symmetry breaking.

\end{quote}

\section{Introduction}

In previous works~\cite{Arb04, Arb05, AVZ, Arb07, AVZ2, Arb09}
N.N. Bogoliubov compensation principle~\cite{Bog1, Bog2, Bog} was applied to studies of spontaneous generation of effective non-local interactions in renormalizable
gauge theories.  Spontaneous generation of Nambu -- Jona-Lasinio like
interaction was studied in works~\cite{Arb05, AVZ, AVZ2} and the description of low-energy
hadron physics in terms of initial QCD parameters turns to be quite successful including values
of parameters: $m_\pi,\, f_\pi,\, m_\sigma,\,
\Gamma_\sigma,\, <\bar q q>, M_\rho, \Gamma_\rho, M_{a_1}, \Gamma_{a_1}$.

In work~\cite{Arb09}  the approach was applied to the electro-weak interaction and a possibility of spontaneous generation of anomalous three-boson interaction of the form
\begin{equation}
-\,\frac{G}{3!}\cdot\,\epsilon_{abc}\,W_{\mu\nu}^a\,W_{\nu\rho}^b\,W_{\rho\mu}^c\,;
\label{FFF}
\end{equation}
was studied. In the present work we continue investigation 
of the electro-weak theory using other approximation scheme, 
which will be formulated in what follows.

The main principle of the approach is to check if an effective interaction
could be generated in a chosen variant of a renormalizable theory. In view
of this one performs "add and subtract" procedure for the effective
interaction with a form-factor. Then one assumes the presence of the
effective interaction in the interaction Lagrangian and the same term with
the opposite sign is assigned to the newly defined free Lagrangian. This
transformation of the initial Lagrangian is evidently identical. However
such free Lagrangian contains completely improper term, corresponding to
the effective interaction of the opposite sign. Then one
has to formulate a compensation equation, which guarantees that this new
free Lagrangian is a genuine free one, that is effects of the
uncommon term sum up to zero. Provided a non-trivial solution of this
equation exists, one can state the generation of the effective
interaction to be possible. Now we apply this procedure to our problem.

In the present work we start with studying of possibility of generation of interaction~(\ref{FFF}).

\section{Compensation equation for anomalous three-boson interaction}\label{alpha}

We start with EW Lagrangian with $3$ lepton $\psi_k$ and colour quark $q_k$ doublets  
with gauge group $SU(2)$. That is we restrict the gauge sector to
triplet of $W^a_\mu$ only. Thus we consider $U(1)$ abelian gauge field $B$ to be decoupled, that means approximation $\sin^2\theta_W \ll 1$.
\begin{eqnarray}
& & L\,=\,\sum_{k=1}^3\biggl(\frac{\imath}{2}
\Bigl(\bar\psi_k\gamma_
\mu\partial_\mu\psi_k\,-\partial_\mu\bar\psi_k
\gamma_\mu\psi
_k\,\biggr)+
\,\frac{g}{2}\,\bar\psi_k\gamma_\mu \tau^a W_\mu^a\psi_k
\biggr)\,+\label{initial}\\
& & +\sum_{k=1}^3\biggl(\frac{\imath}{2}
\Bigl(\bar q_k\gamma_
\mu\partial_\mu q_k\,-\partial_\mu\bar q_k
\gamma_\mu q_k\,\biggr)\,+
\,\frac{g}{2}\,\bar q_k\gamma_\mu \tau^a W_\mu^a q_k
\biggr)\,-\nonumber\\
& &-\frac{1}{4}\,\biggl( W_{\mu\nu}^aW_{\mu\nu}^a\biggr);\qquad
W_{\mu\nu}^a\,=\,
\partial_\mu W_\nu^a - \partial_\nu W_\mu^a\,+g\,\epsilon_{abc}W_\mu^b W_\nu^c\,.
\nonumber
\end{eqnarray}
where we use the standard notations.
In accordance to the Bogoliubov approach~\cite{Bog1, Bog2, Bog} in application to
QFT~\cite{Arb04} we look for
a non-trivial solution of a
compensation equation, which is formulated on the basis
of the Bogoliubov procedure {\bf add -- subtract}. Namely
let us write down the initial expression~(\ref{initial})
in the following form
\begin{eqnarray}
& &L\,=\,L_0\,+\,L_{int}\,;\nonumber\\
& &L_0\,=\,=\,\sum_{k=1}^3\biggl(\frac{\imath}{2}
\Bigl(\bar\psi_k\gamma_
\mu\partial_\mu\psi_k\,-\partial_\mu\bar\psi_k
\gamma_\mu\psi
_k\,\biggr)-\,m_k\bar\psi_k\psi_k\,+\frac{\imath}{2}
\Bigl(\bar q_k\gamma_
\mu\partial_\mu q_k\,-\partial_\mu\bar q_k
\gamma_\mu q_k\,\biggr)-\nonumber\\
& &-\,M_k\bar q_k q_k\biggr)\,-\,\frac{1}{4}\,W_{\mu\nu}^a W_{\mu\nu}^a\,+
\,\frac{G}{3!}\cdot\,\epsilon_{abc}\,W_{\mu\nu}^a\,W_{\nu\rho}^b\,W_{\rho\mu}^c\,;
\label{L0}\\
& &L_{int}\,=\,\frac{g}{2}\,\sum_{k=1}^3\biggl(\bar\psi_k\gamma_\mu
\tau^a W_\mu^a\psi_k\,+\,\bar q_k\gamma_\mu \tau^a W_\mu^a q_k\biggr)\,-\,\frac{G}{3!}\cdot\,\epsilon_{abc}\,
W_{\mu\nu}^a\,W_{\nu\rho}^b\,W_{\rho\mu}^c\,.\label{Lint}
\end{eqnarray}

Here isotopic summation is
performed inside of each quark
bi-linear combination, and notation
 $-\,\frac{G}{3!}\cdot \,\epsilon_{abc}\,
W_{\mu\nu}^a\,W_{\nu\rho}^b\,W_{\rho\mu}^c$ means corresponding
non-local vertex in the momentum space
\begin{eqnarray}
& &(2\pi)^4\,G\,\,\epsilon_{abc}\,(g_{\mu\nu} (q_\rho pk - p_\rho qk)+ g_{\nu\rho}
(k_\mu pq - q_\mu pk)+g_{\rho\mu} (p_\nu qk - k_\nu pq)+\nonumber\\
& &+\,q_\mu k_\nu p_\rho - k_\mu p_\nu q_\rho)\,F(p,q,k)\,
\delta(p+q+k)\,+...;\label{vertex}
\end{eqnarray}
where $F(p,q,k)$ is a form-factor and
$p,\mu, a;\;q,\nu, b;\;k,\rho, c$ are respectfully incoming momenta,
Lorentz indices and weak isotopic indices
of $W$-bosons. We mean also that there are present four-boson, five-boson and
six-boson vertices according to expression for $W_{\mu\nu}^a$
(\ref{initial}).

Effective interaction~(\ref{FFF}) is
usually called {\bf anomalous three-boson interaction} and it is considered for long time on phenomenological grounds~\cite{Hag}. Note, that the first attempt to obtain the anomalous three-boson interaction in the framework of Bogoliubov approach was done in work~\cite{Arb92}. Our interaction constant $G$ is connected with
conventional definitions in the following way
\begin{equation}
G\,=\,-\,\frac{g\,\lambda}{M_W^2}\,.\label{Glam}
\end{equation}
The current limitations for parameter $\lambda$ read~\cite{EW}
\begin{eqnarray}
& &\lambda\,=\,-\,0.016^{+0.021}_{-0.023}\,;\qquad -\,0.059< \lambda < 0.026\,
(95\%\,C.L.)\,.
\label{lambda1}
\end{eqnarray}
Due to our approximation $\sin^2\theta_W\,
\ll\,1$ we use the same $M_W$ for both charged $W^\pm$ and neutral $W^0$ bosons and assume no difference in anomalous interaction for $Z$ and $\gamma$, i.e.
$\lambda_Z\,=\,\lambda_\gamma\,=\,\lambda$.

Let us consider  expression~
(\ref{L0}) as the new {\bf free} Lagrangian $L_0$,
whereas expression~(\ref{Lint}) as the new
{\bf interaction} Lagrangian $L_{int}$. It is important to note, that we
put into the new {\bf free} Lagrangian the full quadratic in $W$ term including
boson self-interaction, because we prefer to maintain gauge invariance of the approximation being used. Indeed, we shall use both quartic term from the last term
in~(\ref{L0}) and triple one from the last but one term of~(\ref{L0}).
Then compensation conditions (see for details~\cite{Arb04}) will
consist in demand of full connected three-gluon vertices of the structure~(\ref{vertex}),
following from Lagrangian $L_0$, to be zero. This demand
gives a non-linear equation for form-factor $F$.

Such equations according to terminology of works
~\cite{Bog1, Bog2, Bog} are called {\bf compensation equations}.
In a study of these equations it is always evident the
existence of a perturbative trivial solution (in our case
$G = 0$), but, in general, a non-perturbative
non-trivial solution may also exist. Just the quest of
a non-trivial solution inspires the main interest in such
problems. One can not succeed in finding an exact
non-trivial solution in a realistic theory, therefore
the goal of a study is a quest of an adequate
approach, the first non-perturbative approximation of
which describes the main features of the problem.
Improvement of a precision of results is to be achieved
by corrections to the initial first approximation.

Thus our task is to formulate the first approximation.
Here the experience acquired in the course of performing
works~\cite{Arb04, Arb05, AVZ, Arb07} could be helpful. Now in view of
obtaining the first approximation we would make the following
assumptions.\\
1) In compensation equation we restrict ourselves by
terms with loop numbers 0, 1.\\
2) We reduce thus obtained non-linear compensation equation to a linear
integral equation. It means that in loop terms only one vertex
contains the form-factor, being defined above, while
other vertices are considered to be point-like. In
diagram form equation for form-factor $F$ is presented
in Fig. 1. Here four-leg vertex correspond to interaction of four
gluons due to our effective three-field interaction. In our approximation we
take here point-like vertex with interaction constant proportional
to $g\,G$.\\
3) We integrate by angular variables of the 4-dimensional Euclidean
space. The necessary rules are presented in paper~\cite{Arb05}.

At first let us present the expression for four-boson vertex
\begin{eqnarray}
& &\frac{V(p,m,\lambda;\,q,n,\sigma;\,k,r,\tau;\,l,s,\pi)}{(2\,\pi)^4} = g G \biggl(\epsilon^{amn}
\epsilon^{ars}\Bigl(U(k,l;\sigma,\tau,\pi,\lambda)-U(k,l;\lambda,\tau,\pi,\sigma)-\nonumber\\& &-U(l,k;\sigma,\pi,\tau,\lambda)+
U(l,k;\lambda,\pi,\tau,\sigma)+U(p,q;\pi,\lambda,\sigma,\tau)-U(p,q;\tau,\lambda,\sigma,\pi)-\nonumber\\
& &-U(q,p;\pi,\sigma,\lambda,\tau)
+U(q,p;\tau,\sigma,\lambda,\pi)\Bigr)-\epsilon^{arn}\,
\epsilon^{ams}\Bigl(U(p,l;\sigma,\lambda,\pi,\tau)-\nonumber\\
& &-U(l,p;\sigma,\pi,\lambda,\tau)
-U(p,l;\tau,\lambda,\pi,\sigma)+
U(l,p;\tau,\pi,\lambda,\sigma)+U(k,q;\pi,\tau,\sigma,\lambda)-\nonumber\\
& &-U(q,k;\pi,\sigma,\tau,\lambda)
-U(k,q;\lambda,\tau,\sigma,\pi)
+U(q,k;\lambda,\sigma,\tau,\pi)\Bigr)+\label{four}\\
& &+\epsilon^{asn}\,
\epsilon^{amr}\Bigl(U(k,p;\sigma,\lambda,\tau,\pi)-U(p,k;\sigma,\tau,\lambda,\pi)
+U(p,k;\pi,\tau,\lambda,\sigma)-\nonumber\\
& &-U(k,p;\pi,\lambda,\tau,\sigma)-U(l,q;\lambda,\pi,\sigma,\tau)
+U(l,q;\tau,\pi,\sigma,\lambda)
-U(q,l;\tau,\sigma,\pi,\lambda)+\nonumber\\
& &+U(q,l;\lambda,\sigma,\pi,\tau)\Bigr)\biggr)\,;\nonumber\\
& &U(k,l;\sigma,\tau,\pi,\lambda)=k_\sigma\,l_\tau\,g_{\pi\lambda}-k_\sigma\,l_\lambda\,g_{\pi\tau}+k_\pi\,l_\lambda\,g_{\sigma\tau}-(kl)g_{\sigma\tau}g_{\pi\lambda}\,.\nonumber
\end{eqnarray}
Here triad $p,\,m,\,\lambda$ {\it etc} means correspondingly incoming momentum, isotopic
index, Lorentz index of a boson.

Let us formulate compensation equations in this
approximation.
For {\bf free} Lagrangian $L_0$ full connected
three-gluon vertices with Lorentz structure~(\ref{vertex}) are to vanish. One can succeed in
obtaining analytic solutions for the following set
of momentum variables (see Fig. 1): left-hand legs
have momenta  $p$ and $-p$, and a right-hand leg
has zero momenta.
However in our approximation we need form-factor $F$ also
for non-zero values of this momentum. We look for a solution
with the following simple dependence on all three variables
\begin{equation}
F(p_1,\,p_2,\,p_3)\,=\,F(\frac{p_1^2\,+\,p_2^2\,+\,p_3^2}{2})\,;\label{123}
\end{equation}
Really, expression~(\ref{123}) is symmetric and it turns to $F(x)$
for $p_3=0,\,p_1^2\,=\,p_2^2\,=\,x$. We consider the representation~(\ref{123})
to be the first approximation and we plan to take into account the
corresponding correction in forthcoming studies.

Now according to the rules being stated above we
obtain the following equation for form-factor $F(x)$
\begin{eqnarray}
& &F(x)\,=\,-\,\frac{G^2\,N}{64\,\pi^2}\Biggl(\int_0^Y\,F(y)\,y dy\,-\,
\frac{1}{12\,x^2}\,\int_0^x\,F(y)\,y^3 dy\,
+\,\frac{1}{6\,x}\,\int_0^x\,F(y)\,
y^2 dy\,+\nonumber\\
& &+\,\frac{x}{6}\,\int_x^Y\,F(y)\,dy\,-\,\frac{x^2}{12}\,
\int_x^Y\,\frac{F(y)}{y}\,dy \Biggr)\,+\,\frac{G\,g\,N}{16\,\pi^2}\,
\int_0^Y\,F(y)\, dy\,+\label{eqF}\\
& &+\,\frac{G\,g\,N}{24\,\pi^2}\,\Biggl(\int_{3 x/4}^{x}\,\frac{(3 x-  4 y)^2 (2 y -3 x)}{x^2 (x-2 y)}F(y)\,
dy\,+\,\int_{x}^Y\,\frac{(5 x- 6 y)}{(x-2 y)}F(y)\, dy\Biggr)\,+\nonumber\\
& &+\,\frac{G\,g\,N}{32 \pi^2}\Biggl(\int_{3 x/4}^{x}\,\frac{3(4 y-3 x)^2(x^2-4 x y+2 y^2)}{8 x^2(2 y-x)^2}\,F(y)\,dy\,+\,\int_x^Y\,\frac{3(x^2-2 y^2)}{8(2 y-x)^2}\,F(y)\,dy
\,+\nonumber\\
& &+\,\int_0^x\frac{5 y^2-12 x y}{16 x^2}\,F(y)\,dy\,+\,\int_x^Y\,
\frac{3 x^2- 4 x y - 6 y^2}{16 y^2}\,F(y)\,dy\Biggr)\,.\nonumber
\end{eqnarray}
Here $x = p^2$ and $y = q^2$, where $q$ is an integration momentum, $N=2$. The last four terms in brackets represent diagrams with one usual gauge vertex (see three last
diagrams at Fig. 1). We introduce here
an effective cut-off $Y$, which bounds a "low-momentum" region where
our non-perturbative effects act
and consider the equation at interval $[0,\, Y]$ under condition
\begin{equation}
F(Y)\,=\,0\,. \label{Y0}
\end{equation}
We shall solve equation~(\ref{eqF}) by iterations. That is we
expand its terms being proportional to $g$ in powers of $x$ and
take at first only constant term. Thus we have
\begin{eqnarray}
& &F_0(x)\,=\,-\,\frac{G^2\,N}{64\,\pi^2}\Biggl(\int_0^Y\,F_0(y)\,y dy\,-\,
\frac{1}{12\,x^2}\,\int_0^x\,F_0(y)\,y^3 dy\,
+\,\frac{1}{6\,x}\,\int_0^x\,F_0(y)\,
y^2 dy\,+\nonumber\\
& &+\,\frac{x}{6}\,\int_x^Y\,F_0(y)\,dy\,-\,\frac{x^2}{12}\,
\int_x^Y\,\frac{F_0(y)}{y}\,dy \Biggr)\,+\,\frac{87\,G\,g\,N}{512\,\pi^2}\,\int_0^Y\,F_0(y)\, dy\,.\label{eqF0}
\end{eqnarray}
Expression~(\ref{eqF0}) provides an equation of the type which were
studied in papers~\cite{Arb04, Arb05, AVZ, Arb07},
where the way of obtaining
solutions of equations analogous to (\ref{eqF0}) are described.
Indeed, by successive differentiation of Eq.(\ref{eqF0}) we come to
Meijer differential equation~\cite{be}
\begin{eqnarray}
& &\biggl(x\,\frac{d}{dx} + 2\biggr)\biggl(x\,\frac{d}{dx} + 1\biggr)\biggl(x\,\frac{d}{dx} - 1\biggr)\biggl(x\,\frac{d}{dx} - 2\biggr)F(x)\,+
\,\frac{G^2\,N\,x^2}{64\,\pi^2}\,F(x)\,=\label{difur}\\
& &=\,4\,\Biggl(-\,\frac{G^2\,N}{64\,\pi^2}\,\int_0^Y F_0(y)
\,y dy + \frac{87\,G\,g\,N}{512\,\pi^2}\,\int_0^Y F_0(y)\, dy
\Biggr)\,;\nonumber
\end{eqnarray}
which solution looks like
\begin{eqnarray}
& &F_0(z) = \,C_1\,G_{04}^{10}\Bigl( z\,|1/2,\,1,\,-1/2,\,-1\Bigr) +
C_2\,G_{04}^{10}\Bigl( z\,|1,\,1/2,\,-1/2,\,-1\Bigr)\,- \label{solution}\\
& &-\,\frac{G\,N}{128\,\pi^2}\,G_{15}^{31}\Bigl( z\,|^0_{1,\,1/2,\,0,\,-1/2,\,-1}\Bigr)\, \int_0^Y \Biggl(G \,y\,-\,\frac{87\, g}{8}\Biggr)F_0(y)\,dy\,
;\nonumber\\
& & G_{15}^{31}\Bigl( z\,|^0_{1,\,1/2,\,0,\,-1/2,\,-1}\Bigr)=
\frac{1}{2\,z}-G_{04}^{30}\Bigl( z\,|1,\,1/2,\,-1,\,-1/2\Bigr)\,;\quad
z\,=\,\frac{G^2\,N\,x^2}{1024\,\pi^2}\,;\nonumber
\end{eqnarray}
where
$$
G_{qp}^{nm}\Bigl( z\,|^{a_1,..., a_q}_{b_1,..., b_p}\Bigr)\,;
$$
is a Meijer function~\cite{be}. In case $q=0$ we write only indices $b_i$ in one
line. Constants $C_1,\,C_2$ are defined by the following boundary conditions
\begin{eqnarray}
& &\Bigl[2\,z^2 \frac{d^3\,F_0(z)}{dz^3}\,+9\,z\,\frac{d^2\,F_0(z)}{dz^2}\,+\,
\frac{d\,F_0(z)}{dz}\Bigr]_{z\,=\,z_0} = 0\,;\nonumber\\
& &\Bigl[2\,z^2\,\frac{d^2\, F_0(z)}{dz^2}\,+5\,z\,\frac{d\, F_0(z)}{dz}\,+\,
F_0(z) \Bigr]_{z\,=\,z_0} = 0\,;
\quad z_0\,=\,\frac{G^2\,N\,Y^2}{1024\,\pi^2}\,.
\label{bc}
\end{eqnarray}

Conditions~(\ref{Y0}, \ref{bc}) defines set of
parameters
\begin{equation}
z_0\,=\,\infty\,; \quad C_1\,=\,0\,
; \quad C_2\,=\,0\,.\label{z0C}
\end{equation}
The normalization condition for form-factor $F(0)=1$ here is the following
\begin{equation}
-\,\frac{G^2\,N}{64\,\pi^2}\,\int_0^\infty F_0(y)
\,y
dy + \frac{87\,G\,g\,N}{512\,\pi^2} \int_0^\infty F_0(y)\,dy\, =\,1\,.
\label{norm}
\end{equation}
However the first integral in (\ref{norm}) diverges due to asymptotics
$$
G_{15}^{31}\Bigl( z\,|^0_{1,\,1/2,\,0,\,-1/2,\,-1}\Bigr)\,\to\,
\frac{1}{2\,z}\,, \quad z\,\to\,\infty\,;
$$
and we have no consistent solution. In view of this we consider the next
approximation. We substitute solution (\ref{solution}) with account of~(\ref{norm}) into terms of Eq.~(\ref{eqF}) being proportional to gauge constant $g$ and
calculate terms  proportional to $\sqrt{z}$. Now we have bearing in mind the normalization condition
\begin{eqnarray}
& &F(z)\,=\,1 + \frac{85\, g\,\sqrt{N} \,\sqrt{z}}{96\,\pi}\Biggl(
\ln\,z + 4\,
\gamma + 4\,\ln\,2 +\frac{1}{2}\,G_{15}^{31}\Bigl( z_0\,|^0_{0, 0, 1/2, -1, -1/2}\Bigr) - \nonumber\\
& &-\,\frac{3160}{357}\Biggr) +
\frac{2}{3\,z} \int_0^z F(t)\,t\, dt + \frac{4}{3\,\sqrt{z}} \int_0^z F(t)
\sqrt{t}\, dt - \frac{4\,\sqrt{z}}{3} \int_z^{z_0} F(t) \frac{dt}{\sqrt{t}}\,+
\nonumber\\
& &+\,\frac{2\,z}{3}\,\int_z^{z_0}\,F(t)\,\frac{dt}{t}\,;\label{eqFg}
\end{eqnarray}
where $\gamma$ is the Euler constant.  
We look for solution of (\ref{eqFg})
in the form
\begin{eqnarray}
& &F(z)\,=\,\frac{1}{2}\,G_{15}^{31}\Bigl( z\,|^0_{1,\,1/2,\,0,\,-1/2,\,-1}
\Bigr) -\,\frac{85\,g \sqrt{N}}{128\,\pi}\,G_{15}^{31}\Bigl( z\,|^{1/2}_{1,\,1/2,
\,1/2,\,-1/2,\,-1}\Bigr)\,+\nonumber\\
& &+\,C_1\,G_{04}^{10}\Bigl( z\,|1/2,\,1,\,-1/2,\,-1\Bigr)\,+
\,C_2\,G_{04}^{10}\Bigl( z\,|1,\,1/2,\,-1/2,\,-1\Bigr)\,.
\label{solutiong}
\end{eqnarray}
We have also conditions
\begin{eqnarray}
& &1\,+\,8\int_0^{z_0}\,F(z)\,dz\,=\,
\frac{87\,g\,\sqrt{N}}{32\,\pi}\,\int_0^{z_0}F_0(z)\,\frac{dz}{\sqrt{z}}\,;\label{g}\\
& &F(z_0)\,=\,0\,;\label{pht1}
\end{eqnarray}
and boundary conditions analogous to~(\ref{bc}). The last
condition~(\ref{pht1}) means smooth transition from the non-trivial
solution to trivial one $G\,=\,0$. Knowing form~(\ref{solutiong}) of
a solution we calculate both sides of relation~(\ref{eqFg}) in two
different points in interval $0\,<\,z\,<\,z_0$ and having four
equations for four parameters solve the set. With $N\,=\,2$ we obtain 
the following solution, which we use to describe the electro-weak case
\begin{equation}
g(z_0)\,=\,0.60366\,;\quad z_0\,=\,9.61750\,;\quad
C_1\,=\,-\,0.035096\,; \quad C_2\,=\,-\,0.051104\,.\label{gY}
\end{equation}
We would draw attention to the fixed value of parameter $z_0$. The solution
exists only for this value~(\ref{gY}) and it plays the role of eigenvalue.
As a matter of fact from the beginning the existence of such eigenvalue is
by no means evident.

Note that there is also solution with a smaller value of $z_0=0.0095531$ and rather large
 $g(z_0)=3.1867$, which with $N = 3$ 
presumably corresponds to strong interaction. This solution
is similar to that considered in work~\cite{Arb07} and it will be studied elsewhere.

We have one-loop expression for $\alpha_s(p^2)$
\begin{equation}
\alpha_{ew}(x)\,=\,\frac{6\,\pi\,\alpha_{ew}(x_0)}{6\,\pi\,+\,5\,\alpha_{ew}(x_0)
\ln(x/x_0)}\,; \quad x=p^2\,; \label{al1}
\end{equation}
We normalize the running coupling
by condition
\begin{equation}
\alpha_{ew}(x_0)\,=\,\frac{g(Y)^2}{4\,\pi}\,=\,0.0290;\label{alphan}
\end{equation}
where
coupling constant $g$ entering in expression
~(\ref{g}) is just corresponding to this normalization point. Note that value~(\ref{alphan}) is not far from physical 
value $\alpha_{ew}(M_W)\,=\,0.0337$. To compare these values 
properly one needs a relation connecting $G$ and $M_W$. For example with $|g\,\lambda|\,=\,0.025,\;\alpha_{ew}(M_W)\,=\,0.0312$. The experimantal value $0.0337$ is reached for $|g\,\lambda|\,=\,0.000003$. For both cases values of $\lambda$ are consistent with limitations~(\ref{lambda1}). Bearing in mind that accuracy of the present approximation is estimted to be $\simeq 10\%$ we can state that agreement is valid for all possible values of $\lambda$. In what follows we shall use experimental value $\alpha_{ew}(M_W)\,=\,0.0337$. 

\section{Four-fermion interaction of heavy quarks}

Let us remind that the adequate description of low-momenta region in QCD can be achieved by an introduction of the effective Nambu -- Jona-Lasinio interaction~\cite{NJL1, NJL2} (see recent review~\cite{RV}). In the framework of the compensation approach the spontaneouis generation of NJL-type interaction was demonstrated in works~\cite{Arb05, AVZ}. In these works pions are described as bound states of light quarks, which are formed due to the effective NJL interaction with account of QCD corrections. 

In the present work we  explore the analogous considerations and assume that scalar fields which substitute elementary Higgs fields are formed by bound states of heavy quarks $t,\,b$. This possibility was proposed in works~\cite{Nambu, Miran, Bardeen} and was considered in a number of publications (see, e.g.~\cite{Lindner}). It comes clear, that estimates of mass of the $t$-quark in this model gives result which exceeds significantly  its measured value. In the present work we obtain the four-fermion interaction in the framework of Bogoliubov compensation approach, while in the previous works on the model the interaction was postulated. In our approach parameters of the problem are obtained as an unique solution of a set of equations quite analogously~\cite{Arb05, AVZ}. In particular we shall see that the $t$-quark mass is quite consistent with the current data.

We have started with Lagrangian~(\ref{initial}) in which both gauge bosons $W$ and spinor particles (leptons and quarks) are massless. As the first stage we consider approximation in which only the most heavy particles aquire masses, namely $W$-s and the $t$-quark while all other ones remain massless. In view of this 
we introduce left doublet $\Psi_L\,=\,(1+\gamma_5)/2 \cdot (t,\,b)$ and right singlet $T_R\,=\,(1-\gamma_5)/2 \cdot t$. Then we study a possibility of spontaneous generation~\cite{Arb04, Arb05, AVZ, AVZ2} 
of the following effective non-local four-fermion interaction
\begin{eqnarray}
& &L_{ff}\,=\,G_1\,\bar \Psi^{\alpha}_L\,T_{R\,\,\alpha}\,\bar T_R^{\beta}\,\Psi_{L\,\beta}+\,G_2\,\bar \Psi^{\alpha}_L\,T_{R\,\,\beta}\,\bar T_R^{\beta}\,\Psi_{L\,\alpha}\,+\nonumber\\
& & \frac{G_3}{2}\,\bar \Psi^{\alpha}_L\,\gamma_\mu\,\Psi_{L\,\,\alpha}\,\bar \Psi_L^{\beta}\,\gamma_\mu\,\Psi_{L\,\beta}+\,\frac{G_4}{2}\,\bar T^{\alpha}_R\,\gamma_\mu\,T_{R\,\,\alpha}\,\bar T_R^{\beta}\,\gamma_\mu\,T_{R\,\beta}\,.\label{ff}
\end{eqnarray}
where $\alpha,\,\beta$ are colour indices. We shall formulate and solve compensation equations for form-factors of the first two interaction, while consideration of the two last ones is postponed for the next approximations. Here we follow the procedure used in works~\cite{AVZ,  AVZ2}, which deal with 
four-fermion Nambu--Jona-Lasinio interaction. However coupling constants $G_3,\,G_4$ essentially influence the forthcoming results.

Following our method (see details in~\cite{AVZ,  AVZ2}) we 
come to the following compensation equations for form-factors 
$F_1(x)$ and $F_2(x), \,x=p^2$, corresponding respectively to the first two terms in~(\ref{ff}). In diagram form the equation is shown at Fig.2.  
\begin{eqnarray}
& &\Phi(x)\,=\,\frac{\Lambda^2(N_c^2 G_1^2+2 N_c G_1 G_2+G_2^2)}{8 \pi^2(N_c G_1+G_2)}\Biggl(1-\frac{N_c G_1+G_2}{8 \pi^2}\int_0^{\bar Y} \Phi(y) dy\Biggr)+\nonumber\\
& &\Biggl(\Lambda^2+\frac{x}{2} \log \frac{x}{\Lambda^2}-\frac{3 x}{4}\Biggr)\frac{G_1^2+G_2^2+2 N_c G_1 G_2+2 \bar G(N_c+1)(G_1+G_2)}{32 \pi^2 (N_c G_1+G_2)}-\nonumber\\
& &\frac{G_1^2+G_2^2+2 N_c G_1 G_2+2 \bar G(N_c+1)(G_1+G_2)}{2^9 \pi^4}\,K\times \Phi\,;\label{PhiF}\\
& &F_2(x)\,=\,\frac{\Lambda^2 G_2}{8 \pi^2}\Biggl(1-\frac{G_2}{8 \pi^2}\int_0^{\bar Y} F_2(y) dy\Biggr)+\nonumber\\
& &\Biggl(\Lambda^2+\frac{x}{2} \log \frac{x}{\Lambda^2}-\frac{3 x}{4}\Biggr)\frac{G_1^2+G_2^2+2 \bar G(G_1+G_2(N_c+1))}{32 \pi^2 G_2}-\nonumber\\
& &\frac{G_1^2+G_2^2+2 \bar G(G_1+G_2(N_c+1)}{2^9 \pi^4}\,K\times F_2\,;\; \Phi({\bar Y})=F_2({\bar Y})=0\,;\nonumber\\
& &\Phi(x)\,=\,\frac{N_c G_1 F_1+G_2 F_2}{N_c G_1+G_2}\,;
\quad \bar G = \frac{G_3+G_4}{2}\,;\quad x=p^2\,;\quad y=q^2\,.\nonumber
\end{eqnarray}
Here $N_c=3$ and a kernel term in equations is the following
\begin{eqnarray}
& &K\times F\,=\,(\Lambda^2-x \log \Lambda^2)\int_0^{\bar Y} F(y) dy-\log \Lambda^2 \int_0^{\bar Y} F(y) y dy+\nonumber\\
& &\frac{1}{6 x}\int_0^x F(y) y^2 dy+\log x \int_0^x F(y) y dy + x(\log x - \frac{3}{2}) \int_0^x F(y) dy+\nonumber\\ 
& &\int_x^{\bar Y} y(\log y-\frac{3}{2})F(y) dy + x \int_x^{\bar Y} \log y F(y) dy + 
\frac{x^2}{6}\int_x^{\bar Y} \frac{F(y)}{y} dy\,.\label{kernel} 
\end{eqnarray}
and $\Lambda$ is auxiliary cut-off, which disappears from all expressions with 
all conditions for solutions be fulfilled.

Introducing substitution $G_1=\rho\,\bar G,\,G_2=\omega \bar G$ and comparing the two equations~(\ref{PhiF}) we get convinced, that both equations become being the same under the following condition
\begin{equation}
\rho=0\,.\label{x}
\end{equation}
and we are rested with  one equation
\begin{eqnarray}
& &F_2(z)\,=\,\frac{\sqrt{\omega^2+8 \omega}}{\omega }\sqrt{z}\,(\log z-3)-16\Biggl[\frac{1}{6\sqrt{z}}\int_0^z F_2(t)
\sqrt{t}\,dt+\nonumber\\
& &\frac{\log z}{2}\int_0^z F_2(t)\,dt\,+\,\frac{\sqrt{z\,}(\log z-3)}{2}\int_0^z\frac{F_2(t)}{\sqrt{t}}dt\,+
\label{eqz}\\
& &\frac{1}{2}\int_z^{\bar z_0} (\log z -3) F_2(t)\,dt\,+\,
\frac{\sqrt {z}}{2}\int_z^{\bar z_0} \log z \frac{F_2(t)}{\sqrt{t}}\,dt\,+\,\frac{z}{6}\int_z^{\bar z_0}\frac{F_2(t)}{t}\,dt
\Biggr]\,;\nonumber\\
& &z\,=\,\frac{(\omega^2+8 \omega)\bar G^2 x^2}{2^{14}\,\pi^4}\,;\quad  t\,=\,\frac{(\omega^2+8 \omega)\bar G^2 y^2}{2^{14}\,\pi^4}\,.\quad \bar z_0\,=\,\frac{(\omega^2+8 \omega)\bar G^2 \bar Y^2}{2^{14}\,\pi^4}\,\nonumber
\end{eqnarray}
Here we omit all terms containing auxiliary cut-off $\Lambda$ 
due to their cancellation.

Performing consecutive differentiations of Eq.(\ref{eqz}) we 
obtain the following differential equation for $F_2$
\begin{equation}
\Biggl(z\frac{d}{dz}+\frac{1}{2}\Biggr)\Biggl(z\frac{d}{dz}\Biggr)\Biggl(z\frac{d}{dz}\Biggr)\Biggl(z\frac{d}{dz}-\frac{1}{2}\Biggr)\Biggl(z\frac{d}{dz}-\frac{1}{2}\Biggr)\Biggl(z\frac{d}{dz}-1\Biggr)\,F_2(z)+\,z\,F_2(z)\,=\,0\,;
\label{diffeq}
\end{equation}
Equation~(\ref{diffeq}) is equivalent to integral equation~(\ref{eqz}) provided the following boundary conditions being 
fulfilled
\begin{eqnarray}
& &\int_0^{\bar z_0}\frac{F_2(t)}{\sqrt{t}}dt\,=\frac{\sqrt{\omega^2+8 \omega}}{8\,\omega }\,;\quad F_2(\bar z_0)\,=\,0\,;\nonumber\\
& &\int_0^{\bar z_0} F_2(t)
\sqrt{t}\,dt\,=\,0\,;\quad 
\int_0^{\bar z_0} F_2(t)\,dt\,=\,0\,.\label{bc1}
\end{eqnarray}
Note that just boundary conditions~(\ref{bc1}) lead to cancellation of all terms containing $\Lambda$. 
Differential equation~(\ref{diffeq}) is a Meijer equation~\cite{be} and the solution of the problem~(\ref{diffeq}, \ref{bc1}) is the following (see also\cite{be} for definition and properties of Meijer functions)
\begin{equation}
F_2(z)\,=\,\frac{1}{2 \sqrt{\pi}} G^{40}_{06}\Bigl(z|0,\frac{1}{2},\frac{1}{2},1,-\frac{1}{2},0\Bigr)\,;\quad \bar z_0\,=\,\infty.\label{sol22}
\end{equation}
Here we also take into account condition $F_2(0)=1$ that gives 
\begin{equation}
\omega=\frac{8}{3}\,.\label{omega}
\end{equation}
We would draw attention to the fact, that unique solution~(\ref{sol22}) exists only for infinite upper limit in integrals. 
\section{Doublet bound state $\bar \Psi_L\,T_R$ }

Let us study a possibility of spin-zero doublet bound state $\bar \Psi_L\,T_R\,=\,\phi$, which can be referred to a Higgs scalar. 
With account of interaction~(\ref{ff}) using results of the previous section we have the following Bethe--Salpeter equation, in which we take into account the $t$-quark mass (see Fig. 3) 
\begin{equation}
\Psi(x)\,=\,\frac{\bar G_2}{16 \pi^2}\int \Psi(y)\,dy\,+\,
\frac{ G^2_2}{2^7\pi^4}\,K^*\times \Psi\,;\label{BS}
\end{equation}
where the modified integral operator $K^*$ is defined in the same way as operator~(\ref{kernel}) with $\bar Y\,=\,\infty$ and lower limit of integration $0$ being changed for $m^2$, 
where $m$ is of order of magnitude of the $t$-quark mass in definition of the kernel (see~\cite{AVZ}).

Then we have again differential equation
\begin{eqnarray}
& &\Biggl(z\frac{d}{dz}-a_1\Biggr)\Biggl(z\frac{d}{dz}-a_2\Biggr)\Biggl(z\frac{d}{dz}\Biggr)\Biggl(z\frac{d}{dz}-\frac{1}{2}\Biggr)\Biggl(z\frac{d}{dz}-\frac{1}{2}\Biggr)\Biggl(z\frac{d}{dz}-1\Biggr)\,\Psi(z)-
\label{diffBS}\\
& &-\,z\,\Psi(z)\,=\,0\,;\; a_1\,=\,-\,\frac{1+\sqrt{1+64\,\mu}}{4}\,;\; a_2\,=\,-\,\frac{1-\sqrt{1+64\,\mu}}{4}\,;\;\mu=\frac{G_2^2\,m^4}{2^{12} \,\pi^4}\,.\nonumber
\end{eqnarray}
where the main difference is the other sign of the last term, 
while variable $z$ is just the same as in~(\ref{diffeq}) with account of relation~(\ref{omega}).
Boundary conditions now are the following
\begin{equation}
\int_\mu^\infty\frac{\Psi(t)}{\sqrt{t}}dt\,=\,0;\;
\int_\mu^\infty \Psi(t)
\sqrt{t}\,dt\,=\,0;\;
\int_\mu^\infty \Psi(t)\,dt\,=\,0;\; \;\Psi(\mu)\,=\,1.\label{bcBS}
\end{equation}
Solution of the problem is presented in the following form

\begin{eqnarray}
& &\Psi(z)\,=\,C_1\,G^{50}_{06}(z|a_1,a_2,\frac{1}{2},\frac{1}{2},1,0)+C_2\,G^{30}_{06}(z|0,\frac{1}{2},1,a_1,a_2,\frac{1}{2})+\nonumber\\
& &C_3\,G^{30}_{06}(z|\frac{1}{2},\frac{1}{2},1,a_1,a_2,0)+C_4\,G^{50}_{06}(z|a_1,a_2,0,\frac{1}{2},1,\frac{1}{2})\,;\label{Psi}
\end{eqnarray}
where $C_i$ for given $\mu$ are uniquely defined by conditions~(\ref{bcBS}).

We define interaction of the doublet $\phi$ with heavy quarks 
\begin{equation}
L_\phi\,=\,g_\phi(\phi^*\bar \Psi_L\,T_R\,+\,\phi\,\bar T_R\,\Psi_L)\,;
\end{equation}
where $g_\phi$ is the coupling constant of the new interaction to be defined by normalization condition of 
the solution of equation~(\ref{BS}). Then we take into 
account the contribution of interaction of quarks with 
gluons and the exchange of $\phi$ as well (see Fig. AA). 
Using standard perturbative method we obtain for the mass 
of the bound state under consideration the following expression in the same way as in~\cite{AVZ}.
\begin{eqnarray}
& &m_\phi^2\,=\,-\,\frac{m^2_t\,I_5}{\sqrt{\pi\,\mu}\,I_2}\,;\quad I_2\,=\,\int_\mu^\infty\frac{\Psi(z)^2\,dz}{z}\,;\label{mphi}\\
& &I_5\,=\,\int_\mu^\infty\frac{(16\,\pi\,\alpha_s(z)-\,g_\phi^2)\,\Psi(z)\,dz}{16\,\pi\,z}\int_\mu^z\frac{\Psi(t)\,dt}{\sqrt{t}}\,.\nonumber
\end{eqnarray}
Here $\alpha_s(z)$ is the strong coupling with standard evolution, normalized at the $t$-quark mass, and we put $m\,=\,m_t$. 
Provided term with brackets inside $I_5$ being positive, bound state $\phi$ is a tachyon.
Let us recall the well-known relation for $t$-quark mass, which is defined by non-zero vacuum average of $(\phi_2^*+\phi_2)/\sqrt{2}$. It reads
\begin{equation}
m_t\,=\,\frac{g_\phi\,\eta}{\sqrt{2}}\,;\label{tusual}
\end{equation}    
where $\eta=246.2\,GeV$ is the value of the electro-weak scalar condensate. However in our approach there are additional contribution to this mass, e.g. due to diagram shown at Fig. 7. That means that for experimental value of the $t$-quark we take the modified definition
\begin{equation}
m_t\,=\,\frac{g_\phi\,\eta}{\sqrt{2}}\,+\Delta M\,=\,\frac{g_\phi\,\eta}{f\,\sqrt{2}}\,.\label{tour}
\end{equation} 
According to these diagrams we have the following expression for $\Delta M$  
\begin{eqnarray}
& &\Delta M\,=\,-\,4\,m_t\,\int_\mu^\infty\frac{F_2(z)\,dz}{\sqrt{z}}\,\int_\mu^\infty\frac{\alpha_s(z)\,F_2(z)\,dz}{2\,\pi\,z}\,-4\,\int_\mu^\infty\frac{m_t(z)\,F_2(z)\,dz}{\sqrt{z}}\,;\label{mt}\\
& &m_t(z)\,=\,m_t\,\biggl(1+\frac{7 \alpha_s(\mu)}{8 \pi}\, \ln \frac{z}{\mu}\biggr)^{-\frac{4}{7}}\,.\nonumber
\end{eqnarray}
Here the first term corresponds to gluon exchange between external legs and the second term corresponds to gluon exchanges inside the loop calculated with account of standard RG mass evolution. Contributions of gluon exchanges from external legs to internal lines cancel. Now parameter $f$ defined in~(\ref{tour}) is the following
\begin{equation}
f=1\,+\,4\,\int_\mu^\infty\frac{F_2(z)\,dz}{\sqrt{z}}\,\int_\mu^\infty\frac{\alpha_s(z)\,F_2(z)\,dz}{2\,\pi\,z}\,+4\,\int_\mu^\infty\frac{m_t(z)\,F_2(z)\,dz}{m_t\,\sqrt{z}}\,.\label{f0f}
\end{equation}
Due to relation~(\ref{bc}) factor $f$ in~(\ref{tour}, \ref{f0f}) is slightly larger than 2.
For strong coupling $\alpha_s(z)$ we use the standard one-loop expression
\begin{equation}
\alpha_s(z)\,=\,\alpha_s(\mu)\biggl(1+\frac{7 \alpha_s(\mu)}{8 \pi}\, \ln \frac{z}{\mu}\biggr)^{-1}\,.\quad \alpha_s(\mu)\,=\,0.108\,;\label{alphas}
\end{equation}
where for strong coupling at the $t$-quark mass we take its value obtained by evolution expression~(\ref{alphas}) from its value at $M_Z$: $\alpha_s(M_Z)\,=\,0.1184 \pm 0.0007$. 
 
Let us consider the possibility when relation~(\ref{mphi}) leads to a tachyon state. For Higgs mechanism to be realized we need also four-fold interaction
\begin{equation}
\L_{\phi4}\,=\,\lambda\,(\phi^*\phi)^2\,.\label{4phi}
\end{equation}
Coupling constant in~(\ref{4phi}) is defined in terms of 
the following loop integral
\begin{equation}
\lambda\,=\,\frac{3\,g_\phi^4}{16\, \pi^2}\,I_4\,;\quad I_4\,=\,\int_\mu^\infty 
\frac{\Psi(z)^4\,dz}{z}\,.\label{lambda}
\end{equation}
From well-known relations $\eta^2\,=\,-m_\phi^2/\lambda$ and  the Higgs mass squared $M_H^2\,=\,-\,2\,m_\phi^2$ we have 
\begin{equation}
\eta^2\,=\,\frac{16 \pi\,m^2_t\,I_5}{3\,g_\phi^4\,\sqrt{\mu}\,I_2\,I_4}\,; \quad M_H^2\,=\,\frac{2\,m^2_t\,I_5}{\pi\,\sqrt{\mu}\,I_2}\,.\label{eta2}
\end{equation}
From~(\ref{tour}) and~(\ref{eta2}) we have useful relation 
\begin{equation}
2\,=\,\frac{16 \pi\,I_5}{3\,g_\phi^2\,f^2\,\sqrt{\mu}\,I_2\,I_4}\,.\label{2}
\end{equation}
We obtain $g_\phi$ from a normalization condition, which is defined by diagrams of Fig. 5
\begin{eqnarray}
& &\frac{3 g^2_\phi}{32 \pi^2}\,\Biggl(I_2\,+\,\frac{\alpha_s(\mu)}{4 \pi}\Bigl(I_{22}^2+2\,I_6\Bigr)\Biggr)\,=\,1;\\\label{normgh}
& &I_{22}\,=\,\int_\mu^\infty\frac{\Psi(t)\,dt}{t}\,;\quad 
I_6\,=\,\int_\mu^\infty\frac{\Psi(z)\,dz}{z\sqrt{z}}\int_\mu^z\frac{\Psi(t)\,dt}{\sqrt{t}}\,.\nonumber
\end{eqnarray}
Here we use strong coupling at the $t$-quark mass~(\ref{alphas}) and perform necessary calculations. In doing this we proceed in the following way: for six parameters 
$\mu,\,g_\phi,\,\eta,\,m_t,\,M_H,\,f$ we have five relations~(\ref{tour}, \ref{eta2}, \ref{2}, \ref{normgh}) and the well-known expression 
\begin{equation}
M_W\,=\,\frac{g_w\,\eta}{2}\,;\label{mw}
\end{equation}
where $g_w$ is weak interaction constant $g$ at $W$ mass. We obtain it by usual RG evolution expression~(\ref{al1}) from value $g$ at $Y$~(\ref{gY}). 
Let us remind that we consider $M_W$ as an input. Thus for the moment we have two input parameters, which are safely known from the experiment
\begin{equation}
M_W\,=\,80.4\,GeV\,;\quad \eta\,=\,246.2\,GeV\,.\label{input}
\end{equation}
The last value corresponds to value of electro-weak coupling 
$g_w(M_W)=0.653$.

Now we present thus obtained parameters 
\begin{eqnarray}
& &\mu\,=\,4.0675\,10^{-12}\,;\quad f\,=\,2.034\,;\quad g_\phi\,=\,2.074\,;\label{parameters}\\
& &m_t\,=\,177.0\,GeV\,;\quad M_H\,=\,1803\,GeV\,.\nonumber
\end{eqnarray}
The most important result here is the $t$-quark mass, which is close to  experimental value $M_t\,=\,173.3\pm 1.1\,GeV$~\cite{MT}. Really, the main difficulty of composite Higgs models~\cite{Nambu, Miran, Bardeen, Lindner} consists in too large $m_t$. Indeed the definition of $g_\phi$ in such models leads to $g_\phi \simeq 3$ and thus $m_t \simeq 500\,GeV$. In the present work we have all parameters, including inportant parameter $f$, being defined by selfconsistent set of equations and the unique solution gives results~(\ref{parameters}), which for $m_t$ is quite satisfactory. The 
large value for $M_H$ seems to contradict to upper limit for this mass, which follows from considerations of Landau pole in the $\lambda \phi^4$ theory. Emphasize, that this limit corresponds to the local theory and in our case of composite scalar fields is not relevant. Such large mass of $H$ means, of course, very large width of $H$
\begin{eqnarray}
& &\Gamma_H\,=\,3784\,GeV\,;\quad BR(H \to W^+\,W^-)\,=\,51.4\%\,;\label{gammah}\\
& &BR(H \to Z\,Z)\,=\,25.6\%\,;\quad BR(H \to \bar t\,t),=\,23.0\%\,.\nonumber
\end{eqnarray}
Thus our approach predicts, that unfortunately quest for Higgs particle at LHC will give negative result. Maybe one could succeed in registration of slight increasing of cross-sections $p+p \to W^+ + W^- +X$, $p+p \to Z + Z +X$, $p+p \to \bar t + t +X$ in region of invariant masses of two heavy particles $1\,TeV < M_{12} < 3\,TeV$.  

For calculations of this section CompHEP package~\cite{comphep} was used.
 
\section{Conclusion}

To conclude we would emphasize, that albeit we discuss quite
unusual effects, we do not deal with something beyond the Standard
Model. We are just in the framework of the Standard Model. What
makes difference with usual results is {\bf non-perturbative non-trivial solution} of
compensation equation. There is of course also {\bf trivial
perturbative solution}. Which of the solutions is realized is to be
defined by stability conditions.  The problem of stability is extremely complicated and needs a special extensive study. 

With the present results we would draw attention to two important achievements provided by the non-trivial non-perturbative solution.
The first one is unique determination of gauge electro-weak coupling
constant $g(M_W)$ in close agreement with experimental
value. The second result consists in calculation of the $t$ - quark mass. At this point we would emphasize, that the
existence of a non-trivial solution itself always leads to
additional conditions for parameters of a problem under study. These two achievements strengthen the confidence in the correctness of
applicability of Bogoliubov compensation approach to the principal problems of elementary particles theory. We consider a check of predictions for Higgs boson mass~(\ref{parameters}) and for its properties~(\ref{gammah}) as a decisive test of validity of the compensation approach.  

\section*{Acknowledgments}

The authors express gratitude to E.E. Boos and V.I Savrin for valuable discussions.

\newpage
\begin{center}
{\bf Figure captions}
\end{center}
\bigskip
\bigskip
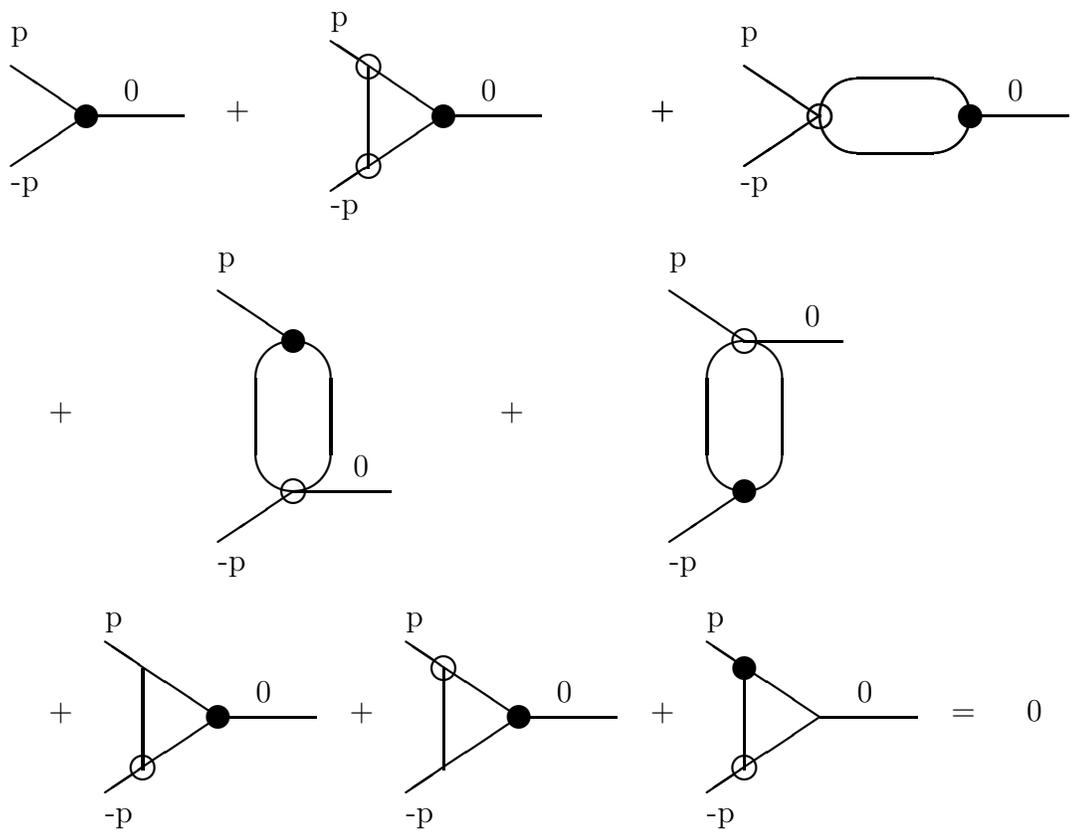
Fig. 1. Diagram representation of the compensation
equation. Black spot corresponds to anomalous three-boson
vertex with a form-factor. Empty circles correspond to point-like anomalous
three-boson and four-boson vertices. Simple point corresponds to usual gauge  vertex.
Incoming momenta are denoted by the corresponding external lines.\\
\\
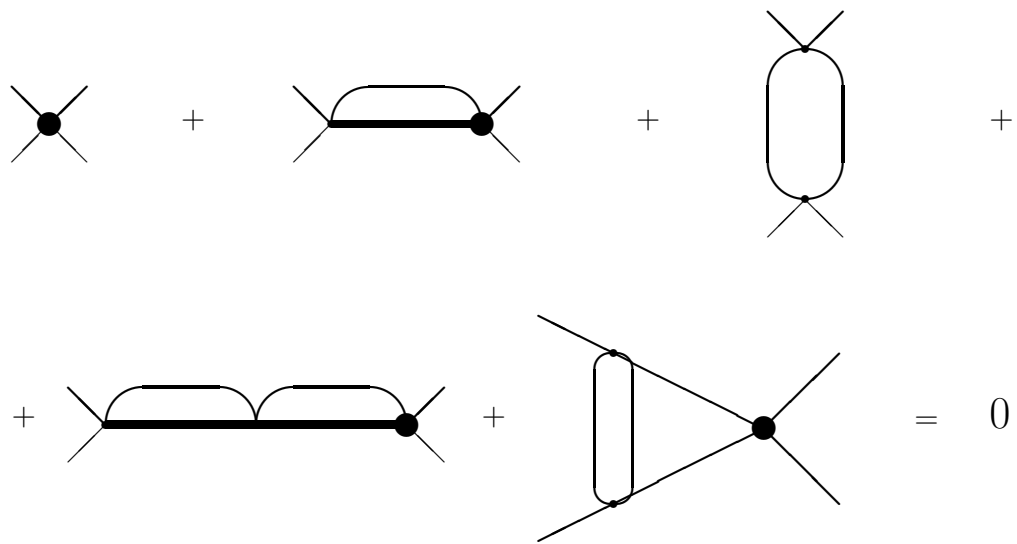
Fig. 2. Diagram representation of the compensation
equation for the four-fermion interaction~(\ref{PhiF}). Lines describe quarks. Simple point
corresponds to the point-like vertex and black circle corresponds to a vertex with a form-factor.\\
\\
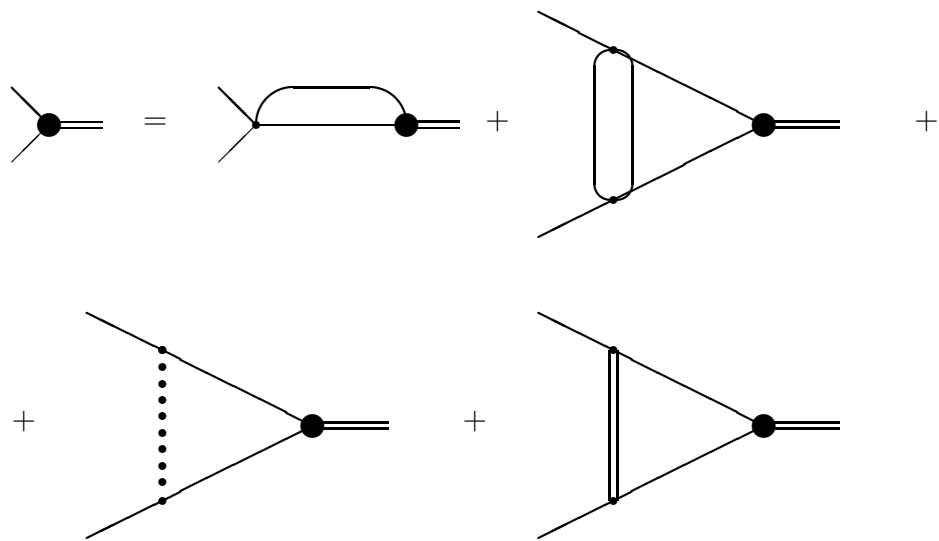
Fig. 3. Diagram representation of the Bethe-Salpeter equation for a bound state of heavy quarks. Double line represent the bound state and dotted line describes a gluon. Black circle corresponds to BS wave function. Other notations are the same as at Fig.2. \\
\\
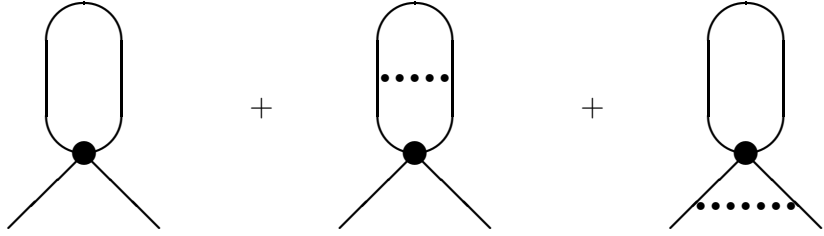
Fig. 4. Diagram representation of additional contribution to the $t$-quark mass. Dotted lines represent gluons. Other notations the same as at Fig. 2.\\
\\
Fig. 5. Diagrams for normalization condition for $H\,\bar \Psi_L\,t_R$-vertex. Notations are the same as at Figs. 2 - 4.\\

\newpage
\begin{picture}(160,105)
{\thicklines
\put(5,90.5){\line(-3,2){10}}
\put(5,90.5){\line(-3,-2){10}}
\put(5,90.5){\circle*{3}}}
\put(5,90.5){\line(1,0){13}}
\put(-5,100.5){p}
\put(-5,80.5){-p}
\put(10,92.5){0}
\put(23.5,90){+}
{\thicklines
\put(52.5,90.5){\line(-3,2){15}}
\put(52.5,90.5){\line(-3,-2){15}}
\put(37.5,102.5){p}
\put(37.5,77.5){-p}
\put(57.5,92.5){0}
\put(52.5,90.5){\circle*{3}}
\put(42.5,83.5){\line(0,1){13.4}}
\put(42.5,83.8){\circle{3}}
\put(42.5,97){\circle{3}}}
\put(52.5,90.5){\line(1,0){13}}}
\put(80,90){+}
{\thicklines
\put(102.5,90.5){\line(-3,2){10}}
\put(102.5,90.5){\line(-3,-2){10}}
\put(102.5,90.5){\circle{3}}
\put(112.5,90.5){\oval(20,10)}
\put(122.5,90.5){\line(1,0){13}}
\put(122.5,90.5){\circle*{3}}
\put(92,100.5){p}
\put(92,80.5){-p}
\put(127.5,92.5){0}
\put(80,90){+}
{\thicklines
\put(102.5,90.5){\line(-3,2){10}}
\put(102.5,90.5){\line(-3,-2){10}}
\put(102.5,90.5){\circle{3}}
\put(112.5,90.5){\oval(20,10)}
\put(122.5,90.5){\line(1,0){13}}
\put(122.5,90.5){\circle*{3}}}
\put(0,50){+}
{\thicklines
\put(32.5,60.5){\line(-3,2){10}}
\put(32.5,40.5){\line(-3,-2){10}}
\put(32.5,50.5){\oval(10,20)}
\put(32.5,40.5){\line(1,0){13}}
\put(32.5,60.5){\circle*{3}}
\put(32.5,40.5){\circle{3}}
\put(22.5,70.5){p}
\put(22.5,30){-p}
\put(40.5,42.5){0}}
\put(60,50){+}
{\thicklines
\put(92.5,40.5){\line(-3,-2){10}}
\put(92.5,60.5){\line(-3,2){10}}
\put(92.5,50.5){\oval(10,20)}
\put(92.5,60.5){\line(1,0){13}}
\put(92.5,40.5){\circle*{3}}
\put(92.5,60.5){\circle{3}}
\put(82.5,70.5){p}
\put(82.5,30){-p}
\put(100.5,62.5){0}}
\put(0,10){+}
{\thicklines
\put(22.5,10.5){\line(-3,2){15}}
\put(22.5,10.5){\line(-3,-2){15}}
\put(7.5,22.5){p}
\put(7.5,-3.5){-p}
\put(27.5,12.5){0}
\put(22.5,10.5){\circle*{3}}
\put(12.5,3.5){\line(0,1){13.4}}
\put(12.5,3.8){\circle{3}}
\put(22.5,10.5){\line(1,0){13}}}
\put(40,10){+}
{\thicklines
\put(62.5,10.5){\line(-3,2){15}}
\put(62.5,10.5){\line(-3,-2){15}}
\put(47.5,22.5){p}
\put(47.5,-3.5){-p}
\put(67.5,12.5){0}
\put(62.5,10.5){\circle*{3}}
\put(52.5,3.5){\line(0,1){13.4}}
\put(52.5,17){\circle{3}}
\put(62.5,10.5){\line(1,0){13}}}
\put(80,10){+}
{\thicklines
\put(102.5,10.5){\line(-3,2){15}}
\put(102.5,10.5){\line(-3,-2){15}}
\put(87.5,22.5){p}
\put(87.5,-3.5){-p}
\put(107.5,12.5){0}
\put(92.5,3.5){\line(0,1){13}}
\put(92.5,3.8){\circle{3}}
\put(92.5,17){\circle*{3}}
\put(102.5,10.5){\line(1,0){13}}}
\put(120,10){=}
\put(130,10){0}
\end{picture}

\bigskip
\bigskip
\bigskip
\bigskip
\bigskip

\begin{center}
Fig. 1.
\end{center}
\newpage
\begin{picture}(160,85)
{\linethickness{1mm}

{\thicklines \put(5,50.5){\line(-1,1){5}}
\put(5,50.5){\line(1,1){5}} \put(5,50.5){\circle*{3}}}
\put(5,50.5){\line(-1,-1){5}} \put(5,50.5){\line(1,-1){5}}

\put(22.5,50){+} {\thicklines \put(42.5,50.5){\line(-1,1){5}}
\put(52.5,50.5) {\oval(20,10)[t]}
\put(62.5,50.5){\line(1,1){5}}\put(42.5,50.5) {\circle*{1}}
\put(62.5,50.5){\circle*{3}}} \put(42.5,50.5){\line(-1,-1){5}}
\put(62.5,50.5) {\line(1,-1){5}} \put(42.5,50.5){\line(1,0){20}}
\put(83,50){+}

{\thicklines \put(105.5,60.5){\line(-1,1){5}}
\put(105.5,60.5){\line(1,1){5}} \put(105.5,50.5){\oval(10,20)}
\put(105.5,60.5){\circle*{1}} \put(105.5,40.5){\circle*{1}}}
\put(105.5,40.5){\line(-1,-1){5}} \put(105.5,40.5){\line(1,-1){5}}
\put(130,50){+} \put(0,10.5){+} {\thicklines
\put(12.5,10.5){\line(-1,1){5}} \put(22.5,10.5) {\oval(20,10)[t]}
\put(12.5,10.5) {\circle*{1}} \put(32.5,10.5){\circle*{1}}}
\put(12.5,10.5){\line(-1,-1){5}} \put(12.5,10.5){\line(1,0){20}}
{\thicklines
 \put(42.5,10.5)
{\oval(20,10)[t]} \put(52.5,10.5){\line(1,1){5}}\put(32.5,10.5)
{\circle*{1}} \put(52.5,10.5){\circle*{3}}}
 \put(52.5,10.5)
{\line(1,-1){5}} \put(32.5,10.5){\line(1,0){20}}
\put(62.5,10.5){+} {\thicklines \put(100,10){\line(-2,1){30}}
\put(100,10){\line(-2,-1){30}} \put(80,10){\oval(5,20)}
\put(80,20){\circle*{1}} \put(80,0){\circle*{1}}
\put(100,10){\line(1,1){10}} \put(100,10){\line(1,-1){10}}
\put(100,10){\circle*{3}}}} \put(120,10){=} \put(130,10){{\Large
0}}
\end{picture}
\ \
\bigskip
\bigskip
\bigskip
\begin{center}
Fig. 2.
\end{center}
\newpage
\begin{picture}(160,55)

{\thicklines \put(5,40.5){\line(-1,1){5}}
\put(5,40.5){\circle*{3}}} \put(5,40.5){\line(-1,-1){5}}
\put(5,40.9){\line(1,0){7}} \put(5,40.1){\line(1,0){7}}
\put(17.5,40){=} {\thicklines \put(32.5,40.5){\line(-1,1){5}}
\put(42.5,40.5){\oval(20,10)[t]} \put(52.5,40.9){\line(1,0){7}}
\put(32.5,40.5) {\circle*{1}} \put(52.5,40.5){\circle*{3}}}
\put(32.5,40.5){\line(-1,-1){5}} \put(52.5,40.1){\line(1,0){7}}
\put(32.5,40.5){\line(1,0){20}} \put(63,40){+} {\thicklines
\put(100,40.5){\line(-2,1){30}} \put(100,40.5){\line(-2,-1){30}}
\put(80,40.5){\oval(5,20)} \put(80,50.5){\circle*{1}}
\put(80,30.5){\circle*{1}} \put(100,40.9){\line(1,0){10}}
\put(100,40.1){\line(1,0){10}} \put(100,40.5){\circle*{3}}}}
\put(120,40){+} \put(0,0){+} {\thicklines
\put(40,0.5){\line(-2,1){30}} \put(40,0.5){\line(-2,-1){30}}
\multiput(20,10.5)(0,-2.2){9}%
{\circle*{1}} \put(20,10.5){\circle*{1}}
\put(20,-9.5){\circle*{1}} \put(40,0.9){\line(1,0){10}}
\put(40,0.1){\line(1,0){10}} \put(40,0.5){\circle*{3}}}
\put(60,0){+} {\thicklines \put(100,0.5){\line(-2,1){30}}
\put(100,0.5){\line(-2,-1){30}} \put(80.5,-9.5){\line(0,1){20}}
\put(79.5,-9.5){\line(0,1){20}} \put(80,10.5){\circle*{1}}
\put(80,-9.5){\circle*{1}} \put(100,0.9){\line(1,0){10}}
\put(100,0.1){\line(1,0){10}} \put(100,0.5){\circle*{3}}
\end{picture}
\bigskip
\bigskip
\bigskip
\bigskip
\bigskip
\bigskip
\bigskip
\bigskip
\bigskip
\bigskip
\bigskip

\begin{center}
Fig. 3.
\end{center}
\newpage
\begin{picture}(160,30)
{\thicklines
\put(33,0){\line(-1,-1){10}}
\put(33,0){\line(1,-1){10}}
\put(33,0){\circle*{3}}
\put(33,10){\oval(10,20)}
\put(77,0){\line(-1,-1){10}}
\put(77,0){\line(1,-1){10}}
\put(77,0){\circle*{3}}
\put(77,10){\oval(10,20)}
\put(73,10){\circle*{1}}
\put(75,10){\circle*{1}}
\put(77,10){\circle*{1}}
\put(79,10){\circle*{1}}
\put(81,10){\circle*{1}}
\put(121,0){\line(-1,-1){10}}
\put(121,0){\line(1,-1){10}}
\put(121,0){\circle*{3}}
\put(121,10){\oval(10,20)}
\put(115,-7){\circle*{1}}
\put(117,-7){\circle*{1}}
\put(119,-7){\circle*{1}}
\put(121,-7){\circle*{1}}
\put(123,-7){\circle*{1}}
\put(125,-7){\circle*{1}}
\put(127,-7){\circle*{1}}
\put(55,5){+}
\put(99,5){+}
}
\end{picture}
\\
\\
\\
\\
\\

\begin{center}
Fig. 4.
\end{center}
\newpage
\newpage
\begin{picture}(140,40)
{\thicklines
\put(0,20){$\textbf{1} \quad \textbf{=}$}
\put(15,20){\line(1,0){15}} \put(15,21){\line(1,0){15}}
\put(55,20){\line(1,0){15}}
\put(42,21){\oval(25,15)}
\put(55,21){\line(1,0){15}}
\put(30,20.5){\circle*{3}}
\put(55,20.5){\circle*{3}}
\put(95,20.5){\circle*{3}}
\put(120,20.5){\circle*{3}}
\put(73,20){$\textbf{+}$}
\put(80,20){\line(1,0){15}} \put(80,21){\line(1,0){15}}
\put(107,27){\circle*{1}}
\put(107,25){\circle*{1}}
\put(107,23){\circle*{1}}
\put(107,21){\circle*{1}}
\put(107,19){\circle*{1}}
\put(107,17){\circle*{1}}
\put(107,15){\circle*{1}} 
\put(120,20){\line(1,0){15}}
\put(107,21){\oval(25,15)}
\put(120,21){\line(1,0){15}}}
  
\end{picture}\qquad\qquad
\\
\\
\\
\\
\\
\begin{center}
Fig. 5.
\end{center}

\begin{thebibliography}{**}
\bibitem{Arb04} B.A. Arbuzov, Theor. Math. Phys., {\bf 140}, 1205 (2004).
\bibitem{Arb05} B.A. Arbuzov, Phys. Atom. Nucl., {\bf 69}, 1588 (2006).
\bibitem{AVZ} B.A. Arbuzov, M.K. Volkov and I.V. Zaitsev, Int. Journ. Mod.
Phys. A, {\bf 21}, 5721 (2006).
\bibitem{Arb07} B.A. Arbuzov, Phys. Lett. B, {\bf 656}, 67 (2007).
\bibitem{AVZ2} B.A. Arbuzov, M.K. Volkov and I.V. Zaitsev, Int. Journ. Mod.
Phys. A {\bf 24}, 2415 (2009).
\bibitem{Arb09} B.A. Arbuzov, Eur. Phys. J. C, {\bf 61}, 51 (2009).
\bibitem{Bog1} N.N. Bogoliubov. Soviet Phys.-Uspekhi, {\bf 67}, 236 (1959).
\bibitem{Bog2} N.N. Bogoliubov. Physica Suppl., {\bf 26}, 1 (1960).
\bibitem{Bog} N.N. Bogoliubov, {\it Quasi-averages in problems of
statistical mechanics.} Preprint JINR D-781, (JINR, Dubna 1961).
\bibitem{Hag} K. Hagiwara, R.D. Peccei, D. Zeppenfeld and K. Hikasa, Nucl. Phys. B, {\bf 282}, 253 (1987).
\bibitem{Arb92} B.A. Arbuzov, Phys. Lett. B, {\bf 288}, 179 (1992).
\bibitem{EW} LEP Electro-weak Working Group, arXiv:
hep-ex/0612034v2 (2006).
\bibitem{be} H. Bateman and A. Erd\'elyi, {\it Higher
transcendental functions. V. 1} (New York, Toronto, London: McGraw-Hill,
1953).
\bibitem{NJL1} Y. Nambu and G. Jona-Lasinio, Phys. Rev. {\bf 122}, 345 (1961).
\bibitem{NJL2} Y. Nambu and G. Jona-Lasinio, Phys. Rev. {\bf 124}, 246 (1961).
\bibitem{RV} M.K. Volkov and A. Radzhabov, Phys. Usp. {\bf 49}, 551 (2006).
\bibitem{Nambu} Y. Nambu, Enrico Fermi Institute Report No 89-08, 1989.
\bibitem{Miran} V.A. Miransky, M. Tanabashi and K. Yamawaki, Phys. Lett B {\bf 221}, 177 (1989).
\bibitem{Bardeen} W.A. Bardeen, C.T. Hill and M. Lindner, Phys. Rev. D {\bf 41}, 1647 (1990).
\bibitem{Lindner} M. Lindner, Int. J. Mod. Phys. A {\bf 8}, 2167 (1993).
\bibitem{MT} Tevatron Electroweak Working Group (CDF and D0 Collaborations), arXiv: 1007.3178 [hep-ex] (2010).
\bibitem{comphep} E.E. Boos et al. (CompHEP Collaboration), Nucl. Instr. Meth. A {\bf 534}, 250 (2004).
\end{thebibliography}
\end{document}